\documentclass[conference]{IEEEtran}
\IEEEoverridecommandlockouts
\usepackage{cite}
\usepackage{amsmath,amssymb,amsfonts}
\usepackage{algorithmic}
\usepackage{graphicx}
\usepackage{subcaption}
\usepackage{makecell}
\usepackage{textcomp}
\usepackage{xcolor}
\usepackage[all]{background}
\usepackage{stackengine}
\setstackEOL{\\}
\setstackgap{L}{\normalbaselineskip}
\SetBgContents{\color{gray}{\tiny \Longstack{PREPRINT - To Appear at ASP-DAC 2023}}}
\SetBgPosition{4.5cm,1cm}
\SetBgOpacity{1.0}
\SetBgAngle{0}
\SetBgScale{1.8}

\def\BibTeX{{\rm B\kern-.05em{\sc i\kern-.025em b}\kern-.08em
    T\kern-.1667em\lower.7ex\hbox{E}\kern-.125emX}}
\begin{document}
\bstctlcite{IEEEexample:BSTcontrol}

\title{Hardware Security Primitives using Passive RRAM Crossbar Array: Novel TRNG and PUF Designs \vspace{-7mm}
}

\author{\IEEEauthorblockN{Simranjeet Singh\IEEEauthorrefmark{1}\textsuperscript{\textsection}, Furqan Zahoor \IEEEauthorrefmark{2}\textsuperscript{\textsection}, Gokulnath Rajendran\IEEEauthorrefmark{2}\textsuperscript{\textsection}, 
Sachin Patkar\IEEEauthorrefmark{1}, \\
Anupam Chattopadhyay\IEEEauthorrefmark{2}, Farhad Merchant\IEEEauthorrefmark{3} \IEEEauthorblockA{\IEEEauthorrefmark{1}Indian Institute of Technology, Bombay,  \IEEEauthorrefmark{2}Nanyang Technological University, Singapore, \IEEEauthorrefmark{3}RWTH Aachen University, Germany}}\{simranjeet, patkar\}@ee.iitb.ac.in, \{anupam@, furqan.zahoor@, gokulnath002@e.\}ntu.edu.sg, merchantf@ice.rwth-aachen.de \vspace{-3mm}}


\maketitle
\begingroup\renewcommand\thefootnote{\textsection}
\footnotetext{Equal contribution}
\endgroup

\begin{abstract}

With rapid advancements in electronic gadgets, the security and privacy aspects of these devices are significant. For the design of secure systems, physical unclonable function (PUF) and true random number generator (TRNG) are critical hardware security primitives for security applications. This paper proposes novel implementations of PUF and TRNGs on the RRAM crossbar structure. Firstly, two techniques to implement the TRNG in the RRAM crossbar are presented based on write-back and 50\% switching probability pulse. The randomness of the proposed TRNGs is evaluated using the NIST test suite. Next, an architecture to implement the PUF in the RRAM crossbar is presented. The initial entropy source for the PUF is used from TRNGs, and challenge-response pairs (CRPs) are collected. The proposed PUF exploits the device variations and sneak-path current to produce unique CRPs. We demonstrate, through extensive experiments, reliability of 100\%, uniqueness of 47.78\%, uniformity of 49.79\%, and bit-aliasing of 48.57\% without any post-processing techniques. Finally, the design is compared with the literature to evaluate its implementation efficiency, which is clearly found to be superior to the state-of-the-art.

\end{abstract}


\section{Introduction}



Physical unclonable functions (PUFs) are being considered the most reliable hardware security primitives especially in IoT era for device authentication and key generation~\cite{liu2016}. PUF usually exploits the variations in the manufacturing process and physical randomness inherent to the device, making them highly reliable and unclonable~\cite{ zhang2018}. PUF utilizes the idea that a device-specific unique key can be derived each time rather than storing a set of keys in the memory. PUF works on the principle of computing a unique output response for the applied input challenge, thus creating a challenge-response pair (CRP), which is crucial for designing security protocols, ranging from device attestation to data encryption. 

Various CMOS-based PUF desingns such as Arbiter PUF~\cite{ganta2013, lim2005} and SRAM PUF~\cite{holcomb2008, guajardo2007} are employed, which depend on the randomness introduced during their fabrication process for their operation. However, they suffer from various demerits, as they are area inefficient~\cite{armknecht2010}, sensitive to temperature variations~\cite{herder2014} and require data post-processing~\cite{rajendran2012}. Also, carbon nanotube-based PUF have been proposed, which utilize the random nature of the bits generated from the unpredictable connections in the CNT devices~\cite{srinivasu2021}. This approach showed some reliability. However, it is not usually employed due to the complexity involved.

\begin{figure}[t!]
    \centering
    \includegraphics[width=\linewidth]{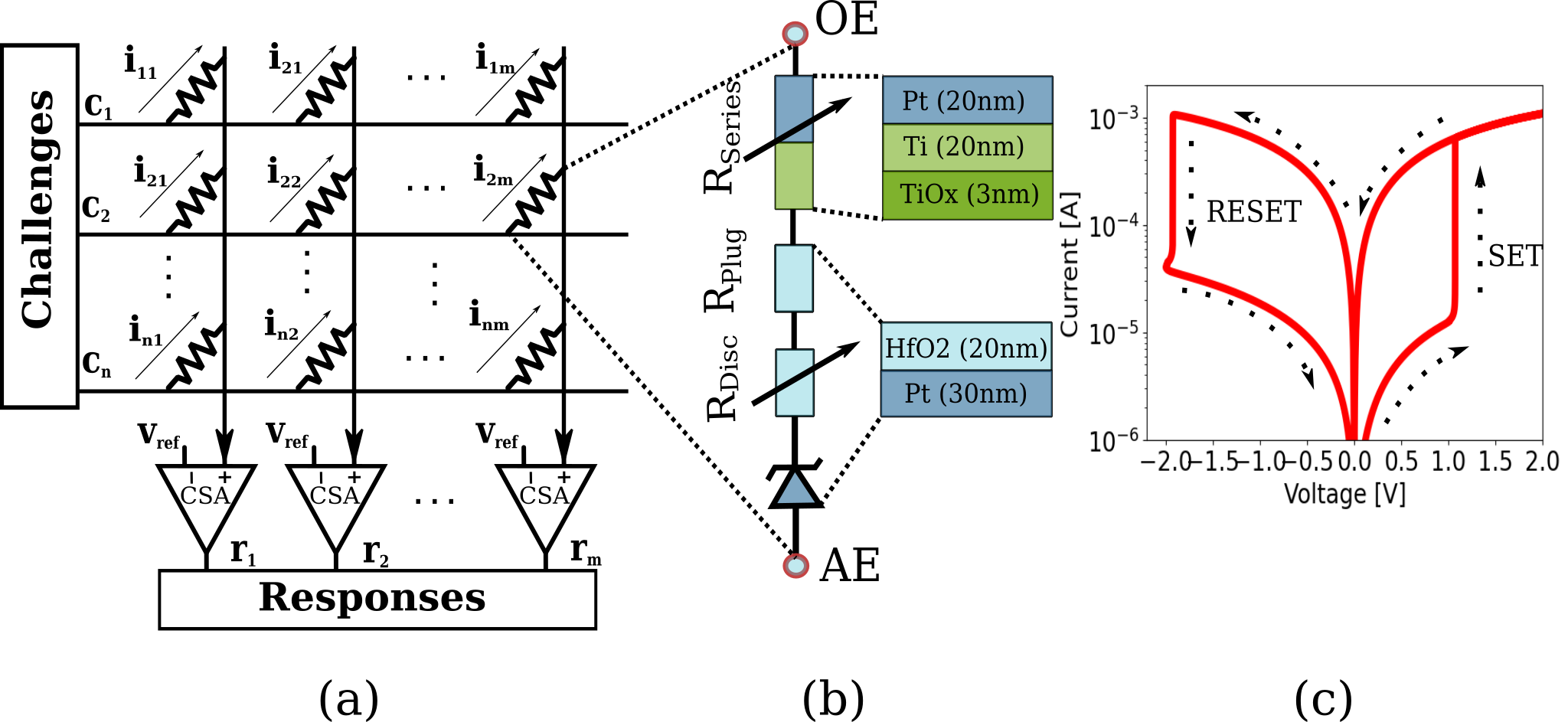}
    \caption{RRAM based PUF design}
    \label{fig:intro}
\end{figure}

In recent years, resistive random access memory (RRAM) has been investigated widely for use as the main memory and in-memory computing. A major drawback to RRAMs being widely adopted as next-generation memories is their stochastic nature and the intrinsic variation in switching parameters~\cite{zahoor2020 }. This uncertain behavior of RRAM is critical for implementing hardware security primitives, including PUF design~\cite{ pang2017}. The schematic structure of RRAM consists of a metal-insulator-metal stack as depicted in Fig.~\ref{fig:intro}b. The RRAM can be considered a variable resistance, where memory states are stored as a high-resistance state (HRS) and a low-resistance state (LRS). The RRAM state is switched by applying a suitable voltage pulse. The devices significantly impact the resistance state of the device due to the device-to-device (D2D) and cycle-to-cycle (C2C) variations~\cite{gonzalez2014}.


This stochasticity of RRAM is desirable for implementing PUF and other hardware security primitives. It utilizes variations during the manufacturing and fabrication process for authentication and secret key storage purposes~\cite{balatti2016}. Lately, multiple research efforts have been focused for utilizing emerging memories such as RRAM~\cite{ gao2022} and STT-MRAM (Spin Transfer Torque Magnetic RAM)~\cite{zhang2014} for various security applications.
 RRAM has emerged as the most feasible option for designs of PUFs since it allows extraction of intrinsic hardware secrets from process variations, similar to that of the traditional silicon-based PUFs. For RRAM, the key is dependent only on the CF variations in the device during LRS and HRS  which are unique for each switching cycle. The secret is independent of the configuration that is selected by a challenge~\cite{gao2020}. These primitives are usually employed for generating unique keys that connect inseparably to the hardware. A typical RRAM-based PUF design is depicted in~Fig.~\ref{fig:intro}. A typical PUF schematic based on RRAM crossbars is depicted along with the crossbar structure for collecting the CRP.

For RRAM, the key is dependent only on the CF variations in the device during LRS and HRS  which are unique for each switching cycle.


In addition to PUFs, TRNGs are another important element for hardware encryption modules. 
TRNGs typically exploit the randomness in physical processes to generate a stream of random numbers~\cite{govindaraj2018}. Although various CMOS-based designs have been implemented, they have limitations as they only provide limited security-specific properties. The emerging technologies, especially RRAM, have shown great potential due to high density, low power operation, and new randomness sources~\cite{yang2021}. The major contributions of this work are as follows:
\begin{itemize}
    \item Two novel TRNG constructions using RRAM crossbar are proposed. The randomness of designed TRNGs is tested using NIST tests.
    \item Investigations on the impact of D2D and C2C variations on the crossbar and how these variations can be exploited to design the PUF are studied. As passive crossbar is used to design the PUF, we exploited sneak-path current as well to construct the PUF.
    \item Lastly, the performance of the proposed 16-bit PUF is evaluated and compared with other RRAM-based PUF designs.
\end{itemize}


The remainder of the paper is organized as follows: Section II details the methods employed to realize the TRNG and PUF designs based on RRAM. Section III presents the results of the RRAM-based security primitives and discusses some critical performance metrics such as uniformity, bit aliasing, and reliability. Section IV concludes the paper.


\section{Methodology}
\label{sec:Methodology}

\begin{figure}
    \centering
    \includegraphics[width=0.9\columnwidth]{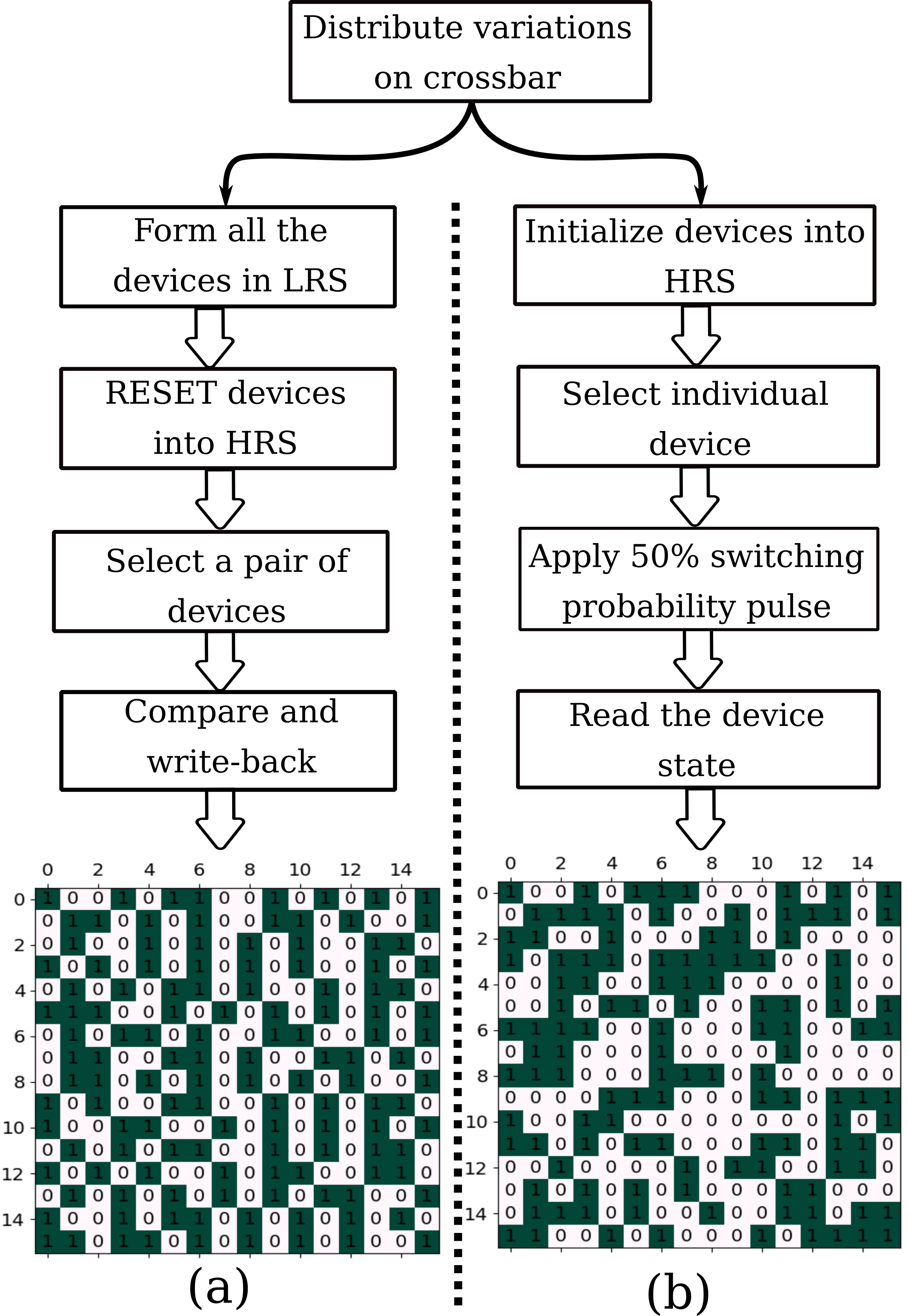}
    \caption{Algorithm to implement TRNG in passive RRAM crossbar using the write-back technique in (a) and 50\% probability pulse in (b).}
     \vspace{-3mm}
    \label{fig:trng}
\end{figure}

Firstly, we demonstrate and validate a novel TRNG based on RRAM that relies on the variations in the material stack of a device in the resistive switching layer. We utilize the inherent stochasticity of the RRAM as the randomness source to achieve the TRNG design. In this work, we propose two variants of TRNG based on RRAM as design elements, as depicted in  Fig.~\ref{fig:trng}. The first design utilizes the write-back technique for generating random numbers in the crossbar, whereas the second design produces random numbers by applying a 50\% switching probability pulse~\cite{john2021}. The proposed RRAM-based TRNG designs have lower circuit complexity advantages as randomness is generated and harvested using simple elements. The proposed TRNG designs can be incorporated into memory subsystems, enhancing security and area efficiency. The random bit streams generated using both designs are depicted in Fig.~\ref{fig:trng}a and Fig.~\ref{fig:trng}b, respectively. 

Fig.~\ref{fig:trng}a shows the flow to implement the TRNG using the write-back technique. Initially, all the devices are formed into the LRS, followed by resetting all devices back to HRS. Next, a pair of devices is selected, and a READ voltage (0.2V) is applied across the rows of selected devices. As the variations on the crossbar are normally distributed, one device will provide more current than another. The current following through these devices are compared, and the fast device in the selected pair is explicitly switched to LRS. This will ensure an equal probability of 0's and 1's in each row of the crossbar, and similar steps keep sliding over the crossbar till the last device.

The second technique characterizes the switching behavior of devices appearing due to D2D and C2C variations. Fig.~\ref{fig:trng}b shows the flow of this technique. Experimentally, a 50\% switching probability pulse is analyzed and applied to the crossbar, which results in 50\% switching in the crossbar from HRS to LRS. Initially, all the devices are in HRS, and Due to the Gaussian distribution of variation on a crossbar, each device will have a different switching threshold. When a 50\% switching probability pulse is applied, random devices in the crossbar switch their state (HRS to LRS), and other remains in the previous state. The switching of devices is a direct function of variation in the crossbar. The bitmap generated using this technique is shown in Fig.~\ref{fig:trng}b. Additionally, to evaluate the randomness of the generated bit streams, the bits generated by the RRAM-based TRNG are passed through the NIST test suite, and the results confirm the random nature of the obtained bit streams. Further, the design TRNGs are used to construct the hardware security primitives.

\begin{figure}
    \centering
    \includegraphics[width=\columnwidth]{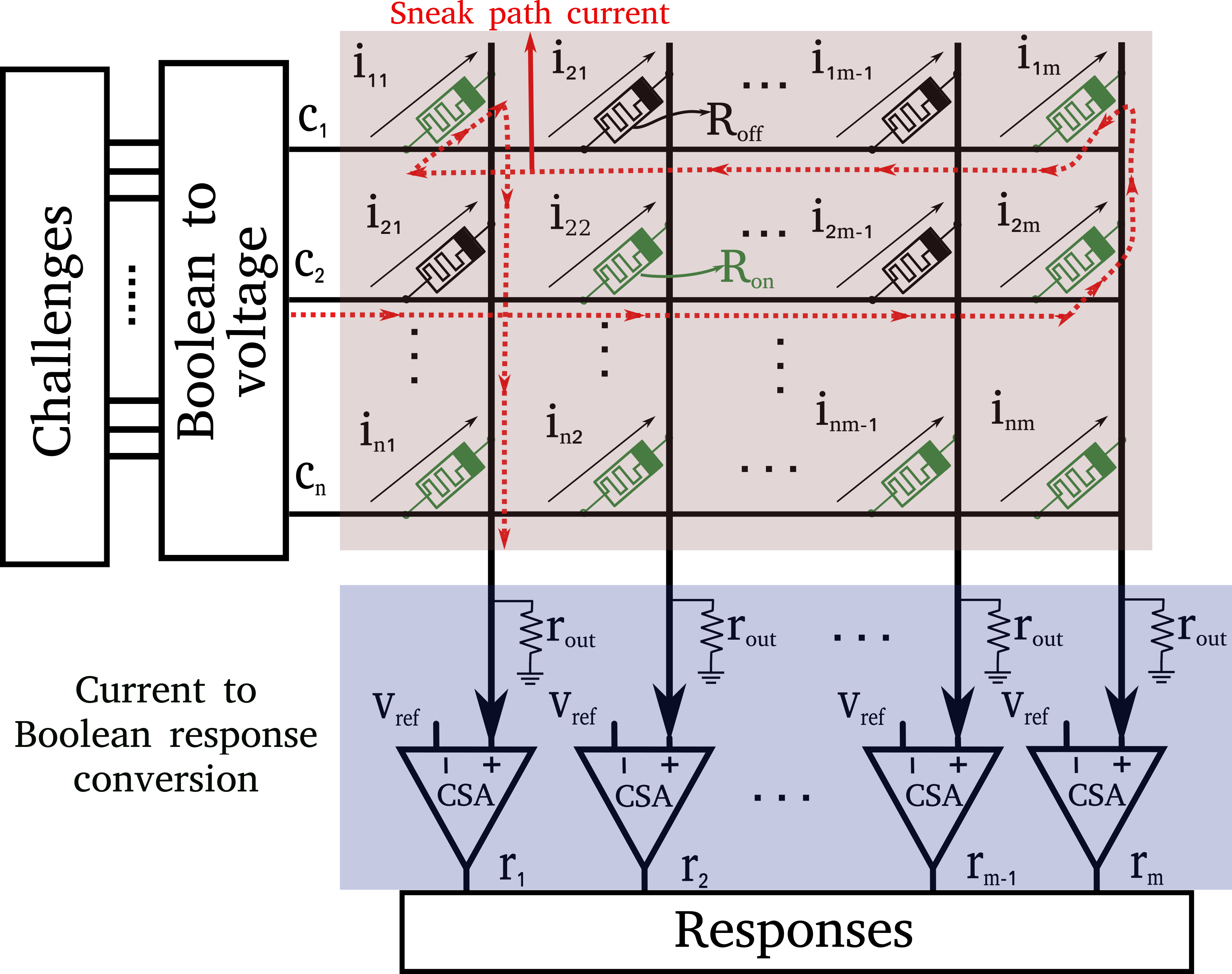}
    \caption{Passive RRAM-based PUF used for challenge-response generation and sneak-path current is marked in red. Where 'n' and 'm' represent the number of rows and columns in the crossbar, respectively. }
    \label{fig:pufcross}
\end{figure}

Next, we propose an architecture to implement the PUF on the passive RRAM crossbar array depicted in Fig.~\ref{fig:pufcross}. The crossbar is first initialized to an entropy source based on the earlier TRNGs algorithm. The resposes of the PUF are collected by applying challenges to the crossbar's rows. These challenges are mapped to a read voltage pulse (1 $\rightarrow$ 0.2V and 0$\rightarrow$ 0V) before applying it to the crossbar. Based on the input challenge and device variations, Kirchoff's current law collects the current following through each device at the column lines. In addition, the sneak-path current in the crossbar contributes to the current at each column, which is entirely random. The current sense amplifier (CSA) converts the analog current values to boolean response bits at the output. Further, many CRPs are collected to check the different PUF properties. The proposed architecture shows an implementation of  TRNG and PUF on a single RRAM crossbar structure.


\section{Experimental Results}
\label{sec:Experimental_results}

This section evaluates the different properties of the proposed PUF on the RRAM crossbar, such as uniformity, uniqueness, bit-aliasing, and reliability. First, we collect the CRPs from the proposed design (refer to Fig.~\ref{fig:intro}), and all mentioned properties are evaluated on this data. We also examine the implementations of different TRNGs in the RRAM crossbar and their validation using the NIST test. Later, we study the impact of D2D and C2C variations on the crossbar and exploitation of the same to construct the TRNG and PUF.

\subsection{RRAM cell Characterization}
The RRAM cell used for this study consists of a Pt/Ti/TiO$_x$/HfO$_2$/Pt material stack, which has been proven to have higher stability in terms of electroforming voltage stability and thermal stability~\cite{Bengel2020 , Hardtdegen2018}. Fig.~\ref{fig:intro}b shows the equivalent device stack of one RRAM cell. The device has two states HRS $\rightarrow$ logic `0' and LRS $\rightarrow$ logic `1'. 
Fig.~\ref{fig:intro}c shows the I-V characteristics of a device where SET (Programming to LRS) and RESET (Programming to HRS) paths are marked with arrows. A voltage pulse with specific duration and amplitude is applied to program the device into HRS or LRS. The device provides resistance between $60 - 100K\Omega$ in HRS and $1.5 - 1.6K\Omega$ in LRS. The devices in the crossbar are used without any selector (Passive) in series. Generally, the passive crossbar faces the problem of sneak-path current, which can be exploited to design the TRNGs and PUF. Next, we examined the impact of process variations on RRAM cells and how these variations can be exploited to design novel hardware security primitives.
\subsection{Variation Analysis}
\label{variations}
In this section, we study the impact of manufacturing process variations on RRAM and how we can exploit these variations to design TRNGs and PUF. There are majorly two types of variations known as D2D and C2C which can affect the switching behavior of RRAM devices. The following are the parameters affected during the manufacturing process of RRAM used in this study:
\begin{itemize}
    \item Oxygen vacancy concentration in the disc, maximum ($N_{Disc,max}$) and minimum ($N_{Disc,min}$).
    \item Radius of the disc ($r_{disc}$).
    \item Length of the disc ($l_{disc}$).
\end{itemize}

Table~\ref{tab:parameters} gives the ranges of variable parameters those are used for this study. These variations are experimentally verified and that's provides a validation to our design choices~\cite{Bengel2020}.


\paragraph{C2C variations} 
C2C variations in the devices are analyzed by changing the variable parameters before every SET and RESET operation. The change of parameters in the next cycle is based on the previous state parameters and random walk of parameters in +ve and -ve directions with an equal probability. The variability parameters are exported in a CSV file using MATLAB scripts. The final parameter file contains variation values and the simulation time stamp at which these parameters will be adopted. This file is then loaded into the \emph{Cadence Spectre} simulation as a parameter list.

\begin{figure}[!t]
    \centering
    \includegraphics[width=\columnwidth]{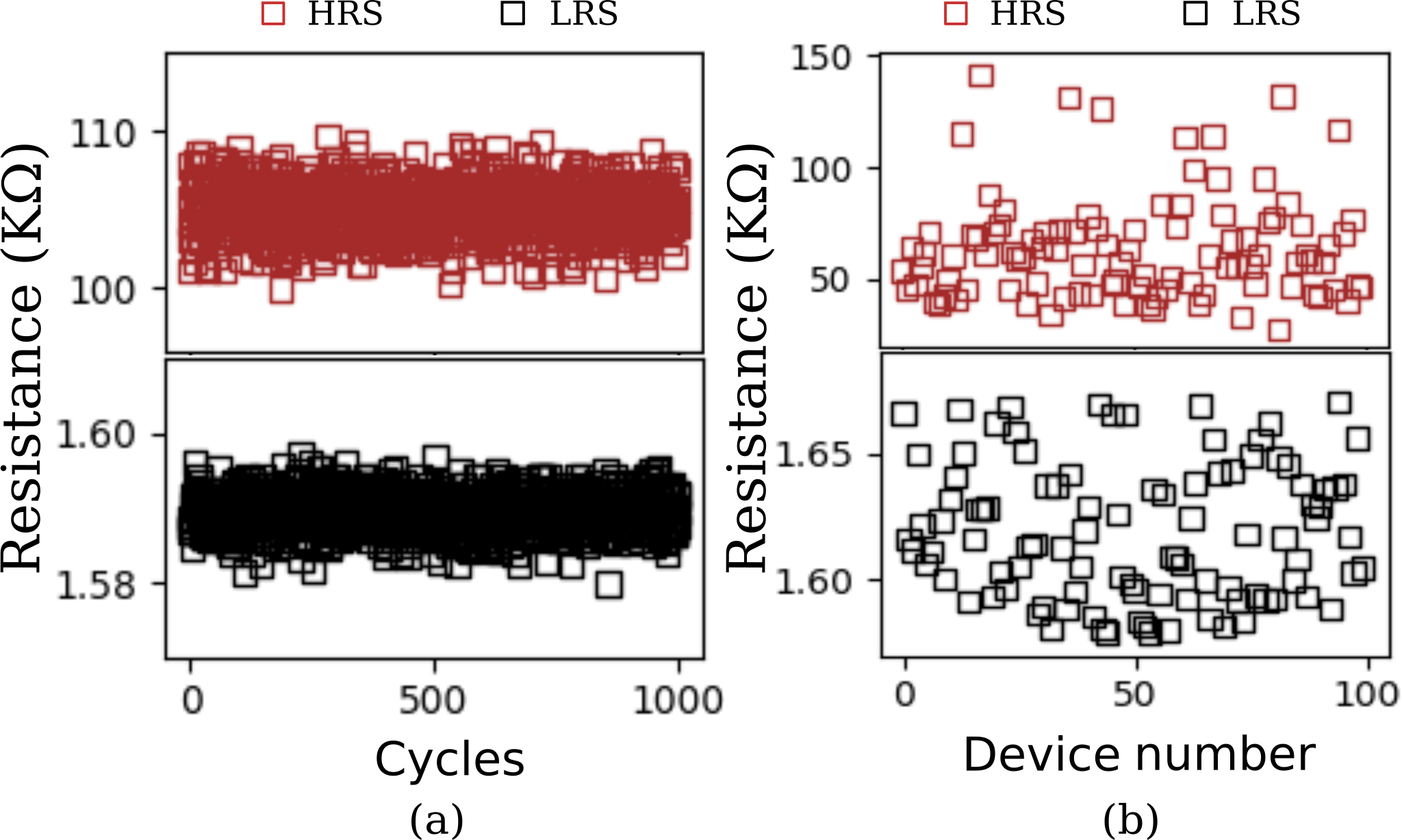}
\caption{LRS and HRS distribution for 1000 cycles under C2C variations in (a) D2D variation on 10x10 crossbar in (b).}
    \label{fig:variations}
\end{figure}

Fig.~\ref{fig:variations}a shows the distribution of  HRS and LRS for 1000 cycles. Each cycle represents a voltage pulse with a 20ns pulse duration, 2.5ns rise time, and 2.5ns fall time. The new set of values are applied after the 100ns period. Within these 100ns, the device is RESET with 25ns pulse of -2V and SET with 2V pulse with the same duration. The state of the device is read by applying a READ pulse of 0.2V after every RESET/SET operation. In the next phase (after 100ns), new values are fetched from the parameter file and applied in the simulation. It has been observed that there is a $\pm$5\% change in HRS and a $\approx$1\% change in LRS during C2C variations.

\begin{table}[b!]
    \centering
    \caption{Parameter ranges for variation study}
    \label{tab:parameters}
\begin{tabular}{|c|c|}
\hline
    \multicolumn{2}{|c|}{RSD = 0.5} \\
    \hline
    - & low / mean / up  \\
    \hline
     $N_{min,var}$ & 4 / 8 / 16  \\
     $N_{max,var}$ & 18 / 20 / 22\\
     $r_{var}$ & 40.5 / 45 / 49.5 \\
     $l_{var} $ & 0.36 / 0.4 /0.44 \\
    \hline
\end{tabular}
    
\end{table}

\paragraph{D2D variations} To simulate D2D variations, the random set of values for each device's variate parameters are drawn from the experimentally verified Gaussian distribution~\cite{Bengel2020}. These variations were then independently applied to the available devices in the crossbar.
Fig.~\ref{fig:variations}b shows the distribution of HRS and LRS states in a 10x10 crossbar (100 devices) without a transistor. 
To measure the HRS distributions, all the devices are first switched to LRS by applying a 2V SET pulse with 150ns pulse width along with 50ns rise and fall time. In the next cycle, the devices are switched to the HRS state by applying RESET voltage of -2V with a similar pulse duration used during SET operation. The resistance state of the device is read at 0.2 V. 
The same procedure is used to find LRS distribution. It has been observed that HRS has more impact of process variations compared to LRS. Due to variations, the HRS
 varies from 31$K\Omega$ to 155$K\Omega$ with an average of 65.56$K\Omega$ and LRS from 1.55$K\Omega$ - 1.67$K\Omega$ with an average of 1.64$K\Omega$. However, HRS has a wide distribution; all HRSs are distinguishable from LRS. A carefully chosen pulse to RESET the devices will lead to switching the devices randomly from LRS to HRS or vice-versa, which can eventually be exploited to design the TRNGs.

\paragraph{Pulse width study}
This study gives the pulse width required to switch a device from LRS to HRS and vice-versa. We analyze the switching characteristic of devices on different pulse duration with a fixed voltage. Table~\ref{tab:probablity} shows the switching probability of crossbar with different pulse duration. To conduct this experiment, all the devices are considered initially into HRS, and a fixed voltage of -0.8V with 1ns rise \& fall time for pulse duration less than 100ns and 10ns for other pulse duration is applied. A total of 100 devices in a 10x10 crossbar are connected to an individual voltage source. A READ voltage of 0.2V is applied to read the state of the devices. The crossbar contains D2D and C2C variations as discussed in Section~\ref{variations}.

\begin{figure}
    \centering
    \includegraphics[width=\columnwidth]{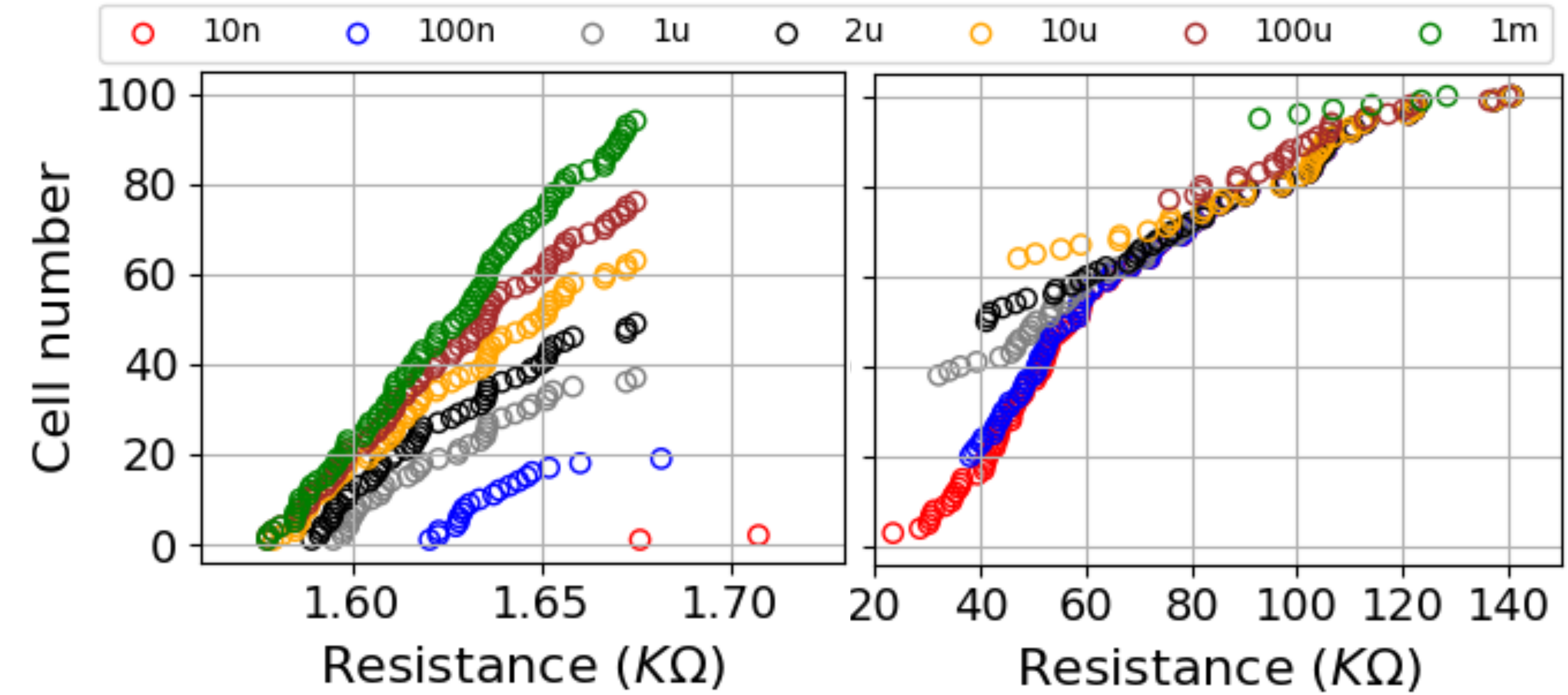}
    \caption{Switching behavior and distribution of LRS and HRS of 100 devices in crossbar under different pulse frequencies.}
    \label{fig:switching}
\end{figure}

Fig.~\ref{fig:switching} shows the switching behavior of devices with different pulse widths. The devices' states are read at 0.2V and sorted according to their resistance value. Initially, all the devices were in HRS. As the pulse width increases, more devices start switching into LRS from the HRS state. This analysis gives a distribution of HRS and LRS, which can further be used to choose the 50\% probability pulse. Finding a pulse width that can randomly switch 50\% of devices in the crossbar is exploited to design the TRNG. A 3us pulse is used for these particular devices to switch the 50\% of devices in the crossbar, as given in Table~\ref{tab:probablity}.


\begin{table}[b!]
    \centering
        \caption{Pulse width study to find out 50\% pulse to generate random numbers in RRAM crossbar. SET voltage = -0.8V, READ voltage = 0.2V, Rise and Fall time = 1ns.}
    \label{tab:probablity}
    \begin{tabular}{|c|c|c|c|}
    \hline
         Pulse duration & SET probability & LRS ($K\Omega$) & HRS ($K\Omega$)  \\
         \hline
         10 ns& 0\% & 1.67 - 1.7 & 23-140.7  \\
         100 ns& 11\% & 1.62-1.67 & 38-140.8  \\
         1 us& 36\% & 1.6-1.675 & 32.2-140.8  \\
        \hline
         \textbf{3 us} & \textbf{50\%} & \textbf{1.58 - 1.675} & \textbf{47- 140.8} \\
         \hline
         10 us& 64\% & 1.585-1.67 & 47-140  \\
         1 ms& 94\% & 1.58-1.67 & 94-128  \\
         10 ms& 100\% & 1.58-1.68 & -  \\

         \hline
         
    \end{tabular}

\end{table}

\subsection{PUF properties}
For evaluating a PUF design, a number of properties need to be analyzed, such as uniformity, uniqueness, bit-aliasing, and reliability. In the following subsections, we provide the analysis of the proposed design on these properties in detail.

\begin{figure*}%
\centering
\begin{subfigure}{0.68\columnwidth}
    \centering
    \includegraphics[width=\columnwidth]{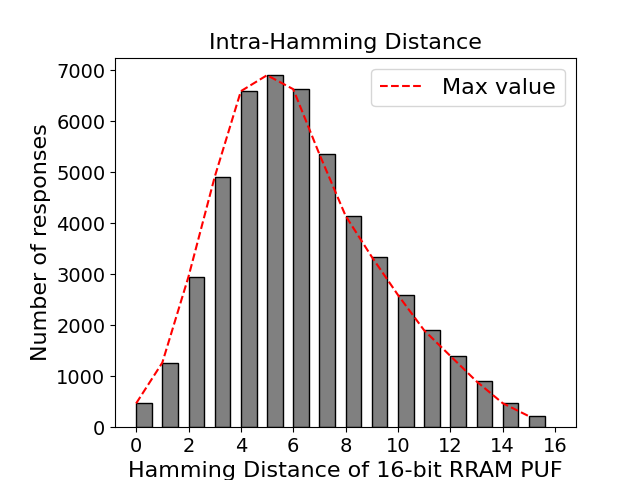}
    \caption{Intra-HD}
    \label{fig:intra_hamm}
\end{subfigure}\hfill%
\begin{subfigure}{.68\columnwidth}
    \centering
    \includegraphics[width=\columnwidth]{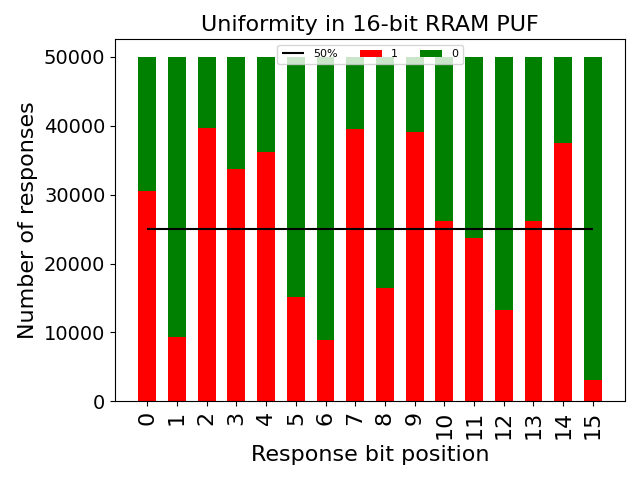}
    \caption{Uniformity }
    \label{fig:uniform}
\end{subfigure}\hfill%
\begin{subfigure}{.68\columnwidth}
    \centering
    \includegraphics[width=\columnwidth]{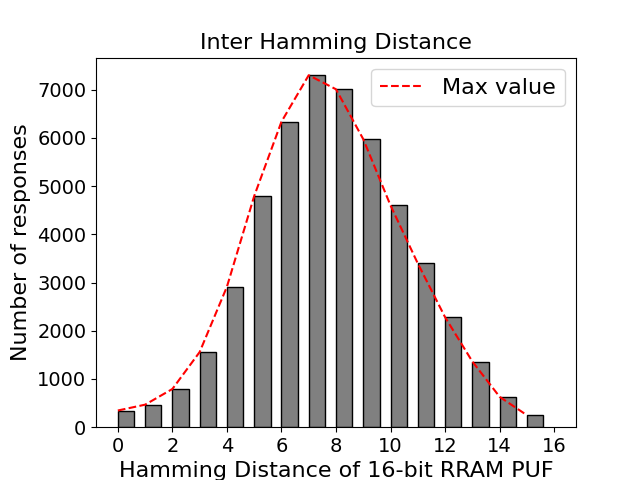}
    \caption{Uniqueness}
    \label{fig:Uniqueness}
\end{subfigure}%
\caption{The different properties of 16-bit RRAM-based PUF}
\vspace{-2mm}
\label{PUF:abc}
\end{figure*}

The first step is calculating the intra Hamming distance (intra-HD), which shows the difference in the output response bits corresponding to different input challenges. Fig.~\ref{fig:intra_hamm} shows the histogram for intra-HD in responses of 16-bit RRAM-based PUF, which is calculated by the Equation~\ref{eq:HD}. Ideally, PUF should result in a 50\% intra-HD in the response bits. In this case, with a particular entropy and variations, the proposed PUF shows 39.68\% intra-HD.

 \begin{equation}
\mbox{\rm Intra-HD} =  \frac{1}{n}\sum_{i=1}^n HD(R_i,R_{i+1}) \times 100
 \label{eq:HD}
 \end{equation}
Where `n' is the total number of challenges given to the PUF, $R_i$ and $R_{i+1}$ are
the responses to the challenges $C_i$ and $C_{i+1}$ respectively. Note that $R_i$ is a bit vector structured from each response bit position at the columns from $r_i$ to $r_m$ for $i_{th}$ challenge.

\paragraph{Uniformity}

The uniformity property shows how the response bit from each device is distributed in 0's and 1's. It is expected from the PUF for an equal number of switching between 0 and 1 for each response bit. The ideal value for uniformity is 50\%, calculated by taking the average of all the response bits' distribution. Fig.~\ref{fig:uniform} shows the distribution of 0's (green) and 1's (red) in each response bit position. Even though some response bits are more inclined towards 0's, and some are towards 1's, the proposed PUF shows 49.79\% uniformity. The distributions of 0's and 1's are a function of the entropy source and will change its inclination on another entropy set for other crossbars.




\paragraph{Uniqueness}
Uniqueness characterizes the ability to distinguish between different devices based on their response to the same challenge.
As the applied challenges are identical, the difference between
the responses is entirely based on the process variations. It is expected that the PUF should respond with 50\% HD in the responses to the same applied challenges for different devices. The uniqueness experiment is conducted using different crossbar variations randomly drawn from the Gaussian distribution. A total of 50K CRPs are collected from each device. Further, Fig.~\ref{fig:Uniqueness} shows the histogram of uniqueness, which is calculated as per Equation~\ref{eq:uniq}. The proposed PUF gives 47.78\% uniqueness, which is very close to the ideal value. 
  \begin{equation}
\mbox{\rm Uniqueness } = \frac{2}{k(k-1)}\sum_{i=1}^{k-1}\sum_{j=i+1}^k\frac{HD(R_i,R_j)}{n} \times 100 
 \label{eq:uniq} 
 \end{equation}

Where, $i$ and $j$ are two different devices. $R_i$ and $R_j$ are the responses from $device_i$ and $device_j$ for the challenge C respectively. $k$ is the number of devices and $n$ is total number of responses.


\paragraph{Bit-aliasing}
Bit-aliasing investigates the switching of each response bit across the number of available crossbars and can be calculated as per Equation~\ref{bitaliasing}. It is essential to ensure that no bit positions of responses are strongly biased towards 0 or 1. To calculate bit-aliasing for the proposed design, CRPs are collected from two different crossbars, and then each bit position of response is analyzed. Ideally, bit-aliasing should be 50\% for a well-balanced design, and the proposed design shows a bit-aliasing of 48.57\%.

  \begin{equation}
\mbox{\rm Bit-aliasing} =  \frac{1}{n}\sum_{i=1}^n (R_{i,p}==1) \times 100
 \label{bitaliasing} 
 \end{equation}
where $R_{i,p}$ is the response bit at the $pth$ cell location in the
$ith$ chip and `n' is the total number of responses including response from all devices.
\paragraph{Reliability}
Reliability captures how efficient a PUF is in repeating the same response for the same challenge. The higher the reliability, the less cost for the error-correcting techniques. Reliability is calculated by intra-HD between responses during the same applied challenges. Ideally, intra-HD for the same response will be zero, and reliability is given as $(100\% - \mbox{\rm (intra-HD)})$. To check the reliability of the proposed design, we applied the same challenge of 10K times and collected the responses. The proposed design shows a reliability of 100\% without any error-correcting techniques.

\subsection{Machine Learning-based Modelling Attacks}
The machine learning (ML) models can be trained to predict the responses of a given PUF. It has been shown that nearly all variants of CMOS-based PUFs can be efficiently modelled using ML~\cite{pufml3} and correspondingly attack resistant PUF structures are also being proposed~\cite{attack_resistant_puf}. These attacks characterize the silicon behavior from CRPs. However, in RRAM-based PUF, the underlined devices have different characteristics and physics compared to the CMOS-based design, which requires a deeper analysis. We reserve the study of detailed analysis of modeling attacks for the future.

\begin{table}[b!]
    \centering
    \setlength{\tabcolsep}{2pt}
    \caption{Comparison of the performance metrics of the proposed PUF}
    \label{tab:comp}
    \begin{tabular}{|c|c|c|c|c|c|c|}
    \hline
    \makecell{\textbf{Parameters}} & \makecell{\textbf{Proposed} \\\textbf{PUF}} &
    \makecell{\textbf{Nano} \\{\textbf{PUF}}~\cite{jey2015} } &
    \makecell{\textbf{RRAM} \\ \textbf{PUF}~\cite{kim2017} } &
    \makecell{\textbf{Memristor} \\ \textbf{PUF}~\cite{Nili2018}} &
    \makecell{\textbf{MR-PUF} \\\cite{Ibrahim2022} }   \\
    \hline
    Crossbar & 16x16 & 1 cell & 64x8 & 2x10x10 (3D) & 1 cell \\
    \hline
        Uniqueness &\textbf{47.78} &47 &49.85 & 50  & 49.3\\
        \hline
        Uniformity & \textbf{49.79}&47 & 47.28 & 49.5 & 49.9\\
        \hline
        Reliability & \textbf{100}&90 & 98.67 & 97 & 100\\
        \hline
        Bit-aliasing & \textbf{48.57}&-  & 49.86 & $\approx$ 50 & 49.6\\
        \hline
    \end{tabular}
    
\end{table}

\subsection{Randomness tests and Comparison}
The process variations in RRAM devices results in different switching behavior of these devices and are exploited to design the TRNG. We proposed two methods (50\% probability pulse and write-back) to generate TRNGs in the 16x16 RRAM crossbar, which are further used to construct the PUF. The national institute of standards and technology (NIST) randomness tests are conducted to analyze the randomness. The method with a 50\% probability pulse passes 11 tests, and the write-back technique passes 08 out of 15 tests. NIST tests are conducted on just 16x16 crossbar data, and it passes all critical tests. The NIST tests can further be improved by increasing the size of the crossbar.

Table~\ref{tab:comp} compares the proposed PUF with other RRAM-based PUFs. Table~\ref{tab:comp} shows that the proposed design on a 16x16 crossbar outperforms other RRAM-based designs, even without the post-processing. MR-PUF~\cite{Ibrahim2022} analyzes the single-cell behavior (not in crossbar) and presents the result after post-processing. Our technique is implemented on the passive crossbar, which exploits the process variations along with the sneak-path current. Construction using a passive crossbar leads to less overall area.



\section{Conclusions}
\label{sec:conclusions}

This work presents novel TRNG and PUF designs based on RRAM technology in the passive crossbar. The inherent variability of the switching mechanism of RRAMs is exploited to generate random bits. The TRNG output was evaluated using the NIST test suite, confirming the generated bits' random nature. Next, the exploitation of device variations and sneak-path currents in the passive RRAM crossbar is shown to implement the PUF. The results are evaluated with extensive SPICE simulations, which verify the robust nature of the proposed PUF  as 100\% reliability is achieved without the need for post-processing. The proposed PUF design also demonstrates good performance for various metrics such as uniformity, uniqueness, and bit-aliasing. The results are encouraging and demonstrate the potential for using RRAM-based designs in hardware security applications. We will explore the prototyping of the proposed designs and a detailed study of resilience against attacks in the future.

\section*{Acknowledgment}

This work was partially funded by Deutsche Forschungsgemeinschaft
(DFG – German Research Foundation) under the priority programme SPP 2253 and was partially funded by the grant NRF-CRP21-2018-0003.

\bibliographystyle{IEEEtran}
\bibliography{Bib}

\end{document}